\documentclass{emulateapj}
\usepackage{natbib}
\usepackage{graphicx}
\usepackage{txfonts}

\newcommand{\gtappeq}{\raisebox{-0.6ex}{$\,\stackrel
{\raisebox{-.2ex}{$\textstyle >$}}{\sim}\,$}}

\begin{document}

\title{Revision of the properties of the GRS 1915+105 jets: Clues from the large-scale structure}

\author{Christian R. Kaiser \& Katherine F. Gunn}
\affil{School of Physics \& Astronomy,University of Southampton, Southampton SO17 1BJ, UK}
\author{Catherine Brocksopp}
\affil{Mullard Space Science Laboratory, University College London, Dorking, Surrey RH5 6NT, UK}
\author{J. L. Sokoloski\footnote{NSF Astronomy and Astropysics Fellow}}
\affil{Harvard-Smithsonian Center for Astrophysics, Cambridge, MA 02138, USA}

\begin{abstract}
The jets of GRS 1915+105 carry a considerable energy away from the central source into the ISM. The similarity of the jets of this source and jets in radio galaxies or radio-loud quasars suggests that we should detect large-scale, synchrotron emitting radio structures surrounding GRS 1915+105.  However, these large structures have not been found. We show that by adapting a model for the radio lobes of extragalatic jet sources we predict a radio surface brightness of the equivalent structures of GRS 1915+105 below the current detection limits. The model uses an energy transport rate of the jets averaged over the jet lifetime. This transport rate is found to be considerably lower than the power of the jets during the rare major ejection events. Thus the lobes contain less energy than would be inferred from these events and produce a lower radio luminosity. The model also predicts a lifetime of the jets of order $10^6$\,years and a gas density of the ISM in the vicinity of GRS 1915+105 of $\sim 150$\,cm$^{-3}$. The impact sites of the jets are identified with two {\sc iras} regions with a flat radio spectrum located on either side of GRS 1915+105. Observations of molecular lines and dust emission from these objects are consistent with our interpretation. Distance estimates for the {\sc iras} regions give 6.5\,kpc and our model implies that this is also the distance to GRS 1915+105. This low distance estimate in combination with the observed motions of jet ejections on small scales yields a jet velocity of about 0.7\,c and an angle of $53^{\circ}$ of the jets to our line of sight. 
\end{abstract}


\section{Introduction}

Collimated outflows or jets are by now a commonly observed feature of X-ray transient binary systems. The similarity of some of the properties of these jets with those of the much larger jets in radio-loud active galaxies has led to the name microquasars for these objects \citep[for a review see][]{mr99}. However, there are also some differences between the jets of microquasars and those of radio galaxies and radio-loud quasars. One of these is the lack of observational evidence for an interaction of the jet flows with their environment in the case of most microquasars. 

Discrete ejections of gas from the central binary system sweep up the gas in front of them. The associated transfer of momentum to this gas should slow the ejections down unless the ejections are much denser than the surrounding gas. However, to date only two examples of decelerating jet ejections have been observed for XTE J1748-288 \citep{hrmshwgp98} and for XTE J1550-564 \citep{cft03}. All other observed jet ejections fade to below the detection limit before any noticeable deceleration has taken place. Although this supports the notion of very dense, ballistic ejections in microquasar jets, such ejecta must also slow down eventually due to the continued loss of momentum to the swept up gas in front of them. 

Another possibility is that the continued ejection of jet material has drilled a channel through the ambient gas. In this case, the jets of microquasars can propagate along this channel without significant deceleration even if the density of the jet material is much lower than that of the surrounding gas. At the end of their channels the jets decelerate violently due to the ram pressure of the ambient gas and the jet material inflates lobes enveloping the jets. This idea was first proposed for the jets of radio-loud active galaxies by \citet{ps74}. The lobes of these objects give rise to a rich morphology of large scale structure detected in radio synchrotron emission \citep[e.g.][]{mg91}. Here we suggest that the jets of microquasars, and those of GRS 1915+105 in particular, give rise to radio lobes emitting synchrotron radiation analogous to those observed in powerful extragalactic radio sources of type FRII \citep{fr74}. In the case of SS433 radio lobes inflated by the jets are indeed observed \citep[e.g.][]{bshma80}.

Searches for a similar structure around the first microquasar identified, GRS 1915+105, have been unsuccessful \citep[hereafter RM98]{rm98}. No radio synchrotron structure comparable to the lobes of radio galaxies has been found. Only two {\sc iras} regions with a flat radio spectra symmetric about GRS 1915+105 and at a distance of several parsec were found. Their flat radio spectra are incompatible with synchrotron emission and they are considerably smaller than the expected radio lobes. One possible explanation for the apparent lack of lobes would be that the jets of GRS 1915+105 do not end in strong shocks which accelerate electrons to relativistic speeds. Only if relativistic electrons are fed into the lobes can we expect to detect radio synchrotron emission. However, it is not clear why the formation of shocks should be suppressed at the end of the jets of GRS 1915+105, since active jets in extragalatic FRII sources always end in strong shocks, the so-called radio hotspots. 

In this paper we develop a self-consistent model of the large scale structure of the jets of GRS 1915+105. We show that by interpreting the two {\sc iras} regions as impact sites of the jets we can successfully apply a fluid dynamical model originally developed for extragalactic jets to this source. The model includes strong shocks at the ends of the jets, but the resulting radio synchrotron emission is too faint to be detected in current radio maps. All other observational results are consistent with our interpretation with the exception of one determination of the distance of GRS 1915+105. However, we discuss this issue in detail in Section \ref{dis}.

In Section \ref{obs} we review the detailed observations of the two {\sc iras} regions by RM98 and \citet[hereafter CRM01]{crm01}. We also derive some basic properties of the gas that gives rise to the detected radiation. Section \ref{model} describes the jet model used and the results obtained from its application to the observations. We discuss the implications of the model for the distance of GRS 1915+105 and for the velocity of the jets in Section \ref{dis}. Finally, Section \ref{conc} summarises our results.

\section{Emission from {\sc iras} 19124+1106 and {\sc iras} 19132+1035}
\label{obs}

In this Section we discuss the observations obtained in various wavebands from the two {\sc iras} sources. We will show in the next Section that all of these observations are consistent with an interpretation of the {\sc iras} sources as sites of excitation of the ISM heated by the strong shocks driven into this material by the GRS 1915+105 jets.

\subsection{Emission from hydrogen}
\label{line}

The flat radio spectrum studied in detail by RM98 (see also Figure \ref{spec}) strongly suggests a bremsstrahlung origin from ionised atomic hydrogen gas with a temperature of at least $10^4$\,K. This view is supported by the detection of the H92$\alpha$ recombination line with a line width of 25\,km\,s$^{-1}$ (RM98). Assuming that the width of the line is caused by thermal Doppler broadening, we would expect a speed of sound of 12.5\,km\,s$^{-1}$ for the emitting hydrogen gas, corresponding to a gas temperature of $T = 1.2 \times 10^4$\,K. Note here that in Section \ref{sage} we will identify the hydrogen gas discussed here with the impact region of the jets from GRS 1915+105. The complexity of the fluid flow in these regions may well contribute to the broadening of the H92$\alpha$ line. The temperature of the gas derived here is therefore an upper limit.

The monochromatic luminosity density of ionised hydrogen due to thermal bremsstrahlung is given by \citep[e.g.][]{ml94}
\begin{equation}
L_{\nu} = 6.8 \times 10^{-50} g \left( \nu, T \right) \frac{n^2}{\sqrt{T}} V \exp \left( -\frac{h \nu}{k_{\rm B} T} \right) \, \mbox{ergs\,s$^{-1}$\,Hz$^{-1}$},
\label{brems}
\end{equation}
where we have assumed that the number density of the electrons, $n$, equals that of the protons and no other ions are present. We also assume that the density and temperature are uniform inside the emitting volume, $V$. The Gaunt factor $g \left( \nu, T \right)$, depends weakly on the temperature and the observing frequency, $\nu$. 

Given the flux density measurements in the radio, we can use equation (\ref{brems}) to estimate the density of the ionised hydrogen gas in the two {\sc iras} sources. For this we assume that all of the emitting gas has the temperature estimated from the width of the H92$\alpha$ line. We do not have any information regarding the extent of the emission regions along our line of sight or their exact three dimensional geometrical structure. Thus it is not straightforward to determine the volume of the shocked ISM. The continuum radio maps of RM98 show angular extents of roughly 35" by 38" for {\sc iras} 19124+1106 and 40" by 33" for {\sc iras} 19132+1035. To account for the uncertainties about the source geometry we express their volume as that of a sphere with a diameter equivalent to 36.5" at the distance of GRS 1915+105 multiplied by an unknown geometrical factor $f$. Given the projected shape of the sources, it is very unlikely that $f$ will be much greater than unity. In principle the emission regions could be hollow shells seen in projection akin to supernova remnants, which would imply $f \ll 1$. However, in this case we would expect the {\sc iras} sources to be edge-brightened which is not the case. Therefore we assume that $f\sim 1$. Note also that the volume as well as most other quantities derived further down will depend on the distance $D$ (measured in kpc) of GRS 1915+105 from us. 

\begin{table*}
\begin{tabular}{cccccccc}
& $T/{\rm K}$ & $V / {\rm pc}^3$ & $\nu / {\rm GHz}$ & $g \left( \nu, T \right)$ & $F_{\nu} / {\rm mJy}$ & $n / {\rm cm}^{-3}$ & $p / 10^{-10}\,{\rm ergs}\,{\rm cm}^{-3}$ \\
\hline
& & & 1.5 & 6.5 & 114 & $650\,f^{-1/2}\,D^{-1/2}$ & $11\,f^{-1/2}\,D^{-1/2}$ \\
19124+1106 & $1.2\times 10^4$ & $0.0029\,D^3\,f$ & 5 & 5.8 & 130 & $740\,f^{-1/2}\,D^{-1/2}$ & $7.1\,f^{-1/2}\,D^{-1/2}$\\
& & & 10 & 5.2 & 114 & $710\,f^{-1/2}\,D^{-1/2}$ & $6.8\,f^{-1/2}\,D^{-1/2}$\\
\hline
& & & 1.5 & 6.5 & 60 & $420\,f^{-1/2}\,D^{-1/2}$ & $4.1\,f^{-1/2}\,D^{-1/2}$\\
19132+1035 & $1.2\times 10^4$ & $0.0029\,D^3\,f$ & 5 & 5.8 & 63 & $510\,f^{-1/2}\,D^{-1/2}$ & $8.4\,f^{-1/2}\,D^{-1/2}$\\
& & & 10 & 5.2 & 52 & $510\,f^{-1/2}\,D^{-1/2}$ & $5.0\,f^{-1/2}\,D^{-1/2}$\\
\hline
\end{tabular}
\caption{Properties of the ionised hydrogen gas observed in thermal bremsstrahlung emission. Radio flux density measurements are from RM98. $D$ is the distance of the emission region from us measured in kpc and $f$ is the volume filling factor with respect to a spherical emission region (see text for details).}
\label{gasprop}
\end{table*}

The results for the density of the ionised hydrogen are summarised in Table \ref{gasprop}. The excellent agreement between the density values derived at the three frequencies for each emission region supports the interpretation of the emission as bremsstrahlung. The densities of the gas in both {\sc iras} sources on either side of GRS 1915+105 are also in remarkably good agreement. The table also lists the pressure in this gas for the assumption of ideal gas conditions.

\subsection{Molecular line emission}
\label{molecular}

CRM01 detected sub-mm line emission due to $^{12}$CO, $^{13}$CO, H$^{13}$CO$^+$ and CS from both {\sc iras}  sources and SiO line emission from {\sc iras}\,19132+1035. The existence of molecules in both {\sc iras} sources alone already demonstrates that some of the gas must be considerably cooler than the ionised atomic hydrogen. Line widths are only available for SiO (4.4\,km\,s$^{-1}$) for {\sc iras}\,19132+1035. For {\sc iras}\,19124+1106 line widths are reported for $^{12}$CO (4.05\,km\,s$^{-1}$), two velocity components of $^{13}$CO (3.88\,km\,s$^{-1}$ and 3.9\,km\,s$^{-1}$) and two components of CS (3.039\,km\,s$^{-1}$ and 1.589\,km\,s$^{-1}$). If these line widths are due to thermal Doppler broadening, then they imply gas temperatures of a few to several 100s of Kelvin, considerably cooler than the atomic hydrogen gas. Both CS and SiO emission are considered to indicate high gas densities comparable to the central parts of H{\sc ii} regions \citep[e.g.][]{rw04}. The low temperatures and high density of the molecular gas probably indicate that the {\sc iras} regions contain a mixture of hydrogen gas with embedded denser clouds of molecular material. It is also interesting to note that the shock tracer SiO is only detected in {\sc iras} 19132+1035 but not in {\sc iras} 19124+1106. However, it is not clear from the data presented in CRM01 whether the molecular material is indeed co-spatial with the ionised hydrogen gas.

\subsection{Synchrotron emission from gas}
\label{nontherm}

The radio maps of {\sc iras} 19132+1035 show a feature with a steep radio spectrum to the northwest of the main flat spectrum region (RM98). This feature is elongated with a length of 17.4" and a width of 7.9". The spectrum is consistent with a power law $F_{\nu} \propto \nu^{\alpha}$ with $\alpha = -0.8$. The spectral index is typical for synchrotron emission and therefore implies the presence in this feature of a relativistic plasma consisting of magnetic field and a population of relativistic electrons. We can use the standard minimum energy argument to estimate the properties of this relativistic plasma \citep[e.g.][]{ml94}. Here we neglect any possible contribution to the energy content of the emission region by protons. For an emission region of monochromatic luminosity $L_{\nu}$ and volume $V$ the strength of the magnetic field is given by
\begin{equation}
B_{\rm min} = \left[ \frac{3}{2} G\left( \alpha \right) \frac{L_{\nu}}{V} \right]^{2/7}.
\label{bmin}
\end{equation}
The constant $G\left(\alpha \right)$ depends on the observing frequency $\nu$, the spectral index $\alpha$ and to a lesser extent on the cut-offs of the synchrotron spectrum, $\nu_{\rm min}$ and $\nu_{\rm max}$. We do not know over which range in frequency the synchrotron spectrum extends. We therefore arbitrarily set $\nu_{\rm min} = 10^7$\,Hz and $\nu_{\rm max} =10^{11}$\,Hz. Changing either of these parameters does not change our results significantly. 

\begin{table*}
\begin{tabular}{cccc}
$F_{1.5}/{\rm mJy}$ & $V/{\rm pc}^3$ & $B_{\rm min} / \mu{\rm G}$ & $p / 10^{-10}\,{\rm ergs}\,{\rm cm}^{-3}$ \\ 
\hline
5 & $9.9\times 10^{-5}\,D^3\,\left( \sin \theta \right)^{-1}$ & $120\,\left( D^{-1} \sin \theta \right)^{2/7}$ & $4.3\,\left( D^{-1} \sin \theta \right)^{4/7}\,\left( k+1 \right)$ \\
\hline
\end{tabular}
\caption{Properties of the non-thermal feature next to {\sc iras} 19132+1035. Radio flux density measurement at $1.5$\,GHz from RM98. $D$ is the distance of the emission region from us measured in kpc, $\theta$ is the angle of the jet leading to {\sc iras} 19132+1035 to our line of sight and $k$ is the ratio of the energy of non-radiating particles and that of the relativistic plasma (see text for details).}
\label{nthprop}
\end{table*}

The radio observations of the non-thermal feature imply a luminosity density of $L_{\nu} = 6.0 \times 10^{18}\,D^2$\,ergs\,s$^{-1}$\,Hz$^{-1}$ at 1.5\,GHz. We assume a cylindrical geometry for the feature with a radius equivalent to 3.9" at the distance of GRS 1915+105 and a length of 17.4"\,$\left( \sin \theta \right)^{-1}$, where $\theta$ is the angle of the long axis of the cylinder to our line of sight. Using equation (\ref{bmin}) we find the strength of the magnetic field, $B_{\rm min}$. For minimum energy conditions the total pressure in this feature is then given by 
\begin{equation}
p = \frac{7}{9}\,\frac{B_{\rm min}^2}{8 \pi} \left( k +1 \right) ,
\label{minpres}
\end{equation}
where $k$ is the ratio of the internal energy stored in any particles which do not contribute to the emission of synchrotron radiation to the sum of the energy in the magnetic field and in the relativistic electrons. All results are summarised in Table \ref{nthprop}.

\subsection{Dust}

\begin{figure}
\centerline{
\includegraphics[width=10cm]{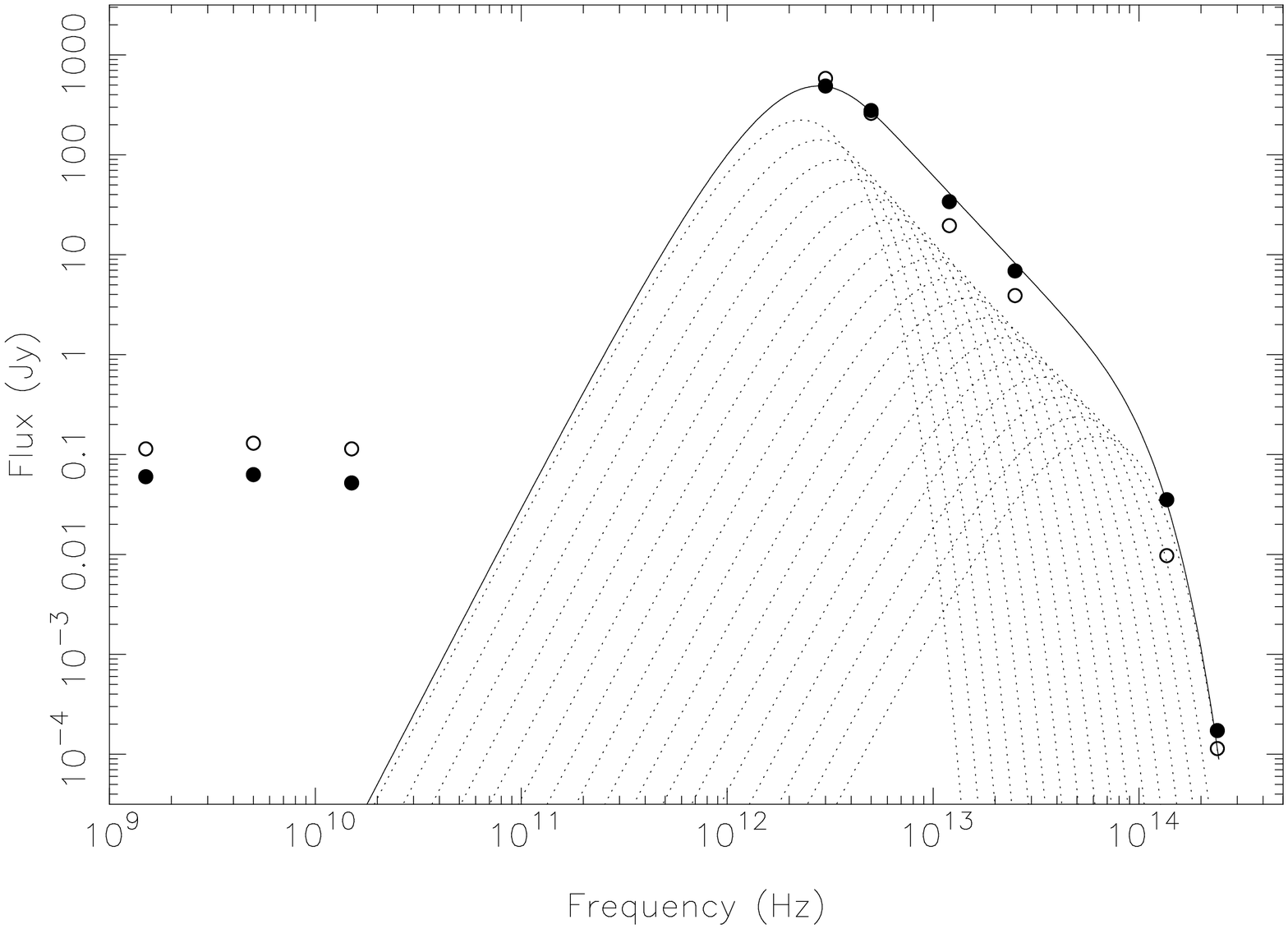}}
\caption{Continuum emission spectrum of the two {\sc iras} regions from radio to near infrared wavelengths. The observational data (symbols) was taken from CRM01: Closed symbols: {\sc iras} 19124+1106, open symbols: {\sc iras} 19132+1035. The lines show our model for the IR emission  of {\sc iras} 19124+1106 by dust with a range of temperatures (see text for details). The dotted lines show the individual contributions to the emission from dust with a single temperature. The solid line shows the total emission from all dust components.}
\label{spec}
\end{figure}

The two IRAS sources detected by CRM01 (see their Table 2) show IR emission with a spectral energy distribution of the form $S_\nu \propto \nu^{-\alpha}$ at $\nu > 3\times10^{12}$ Hz ($\lambda < 100
\mu$m) with $\alpha < 3$.  Since the high frequency cutoff in a thermal dust spectrum is much steeper than this, we infer a range of dust temperatures.  The peak of the dust emission must lie at wavelengths longer than those observed by CRM01, i e. $>100\mu$m, and therefore the bulk of the dust
has a temperature of 80K or lower.

Dust emission is usually modelled using a modified black-body spectrum, $B(\nu, T_{\rm d})$, of the
form:
\begin{equation}
L_\nu = 4\pi \kappa_\nu B (\nu,T_{\rm d}) M_{\rm d}.
\label{dusteq}
\end{equation}
The opacity is parametrized as $\kappa_\nu \propto
\nu^\beta$ where $\beta$ usually lies in the range $1<\beta<2$.  This
relationship is normalised such that at $850\mu$m (353\,GHz),
$\kappa_\nu = 0.77 \, {\rm cm}^2 \, {\rm g}^{-1}$ \citep{SLUGS1}.
Here we take the value $\beta=2$, which gave the best fit to the
observed properties of nearby galaxies \citep{SLUGS2}, and can
therefore be used as a good approximation to the average properties of
the interstellar medium in normal galaxies.  This value of $\beta$ is
consistent with measurements of different environments within our own
Galaxy \citep[see e.g.][]{Gordon88,Sodroski97,Ristorcelli98}.

We model the dust mass
distribution as a powerlaw $dM = C_{\rm dust}  \left( T / T1 \right)^{-\alpha} dT$, within the
range $T1 < T_{\rm dust} < T2$.  We take $T1 = 20$K, which is the
average cold dust temperature of nearby galaxies \citep{SLUGS1}.  The
near IR $J$-- and $K$--band observations (at 1.1 and 1.6$\mu$m)
provide a clear cutoff in the emission spectrum, giving a maximum
temperature $T2 = 600$K for {\sc iras} 19124+1106 and $T2=750$K for {\sc iras} 19132+1035. The highest temperature components however
contribute the least to both the total mass and the total energy
output of the system.  In this way, we find good agreement with the
IR data (see Table \ref{dustprop} and Figure \ref{spec}).

\begin{table*}
\begin{tabular}{cccccc}
& $T2/K$ & $\alpha$ & $L / 10^{36} {\rm ergs}\,{\rm s}^{-1}$ & $M_{\rm tot} / M_{\odot}$ & $C_{\rm dust} / M_{\odot}\, K^{-1}$\\
\hline
19124+1106 & 600 & 7.2 & $3.1\,D^2$ & $0.38\,D^2$ & $0.12\,D^2$\\
19132+1035 & 750 & 7.8 & $3.5\,D^2$ & $0.56\,D^2$ & $0.19\, D^2$\\
\hline
\end{tabular}
\caption{Properties of the dust associated with the {\sc iras} emission regions. $D$ is the distance of the emission regions from us measured in kpc.}
\label{dustprop}
\end{table*}

\section{The jet model}
\label{model}

\subsection{Basic setup}
\label{basic}

For interpreting the interactions of the large-scale jet structure of GRS 1915+105 with its environment we will use the model developed in \citet[hereafter KA97]{ka96b} for the jets of radio galaxies and radio-loud quasars. The model is based on twin jets emerging from the central source in opposite directions. The supersonic jets end in strong shock fronts at the location where they impact on the ambient gas. The gas coming up the jets inflate two lobes which are overpressured with respect to the environment. The lobes therefore expand sideways and thus a bow shock is driven into the surrounding ISM. The pressure in the lobes confines the jets which are propagating through this region. By inflating the lobes the jets create their own high pressure but low density environment protecting them from disruption due to turbulent gas entrainment. This ensures that the jets do not widen much on their way through the lobes and thus the momentum and energy transported by the jets are delivered to the jet shocks without significant losses. 

\begin{figure}
\centerline{
\includegraphics[width=8.45cm]{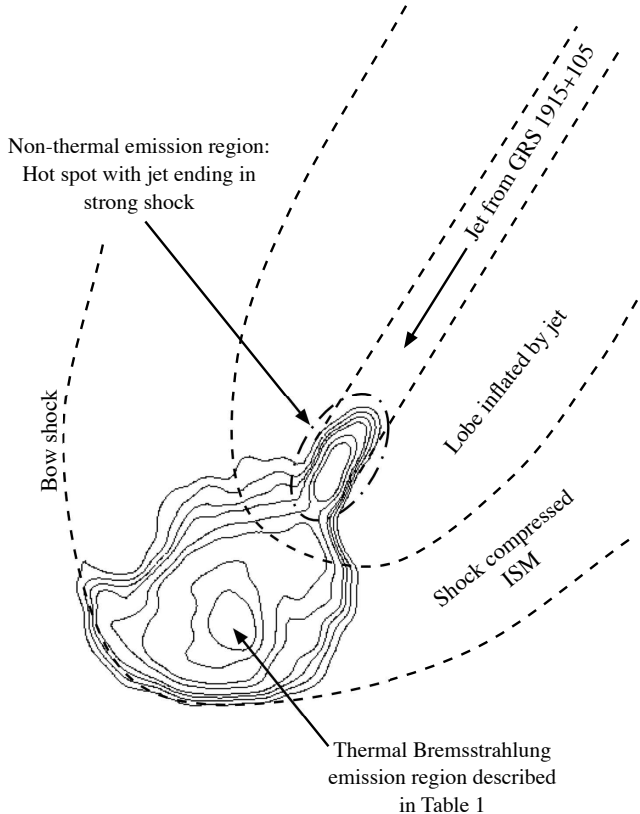}}
\caption{Sketch of the model setup overlayed on the 20\,cm radio contours of {\sc iras} 19132+1035 by RM98. Dashed lines show model features not directly observed in the radio.}
\label{sketch}
\end{figure}

Here we identify the two {\sc iras} sources discussed above with the impact sites of the two jets of GRS 1915+105. The basic setup is sketched in  Figure \ref{sketch} for {\sc iras} 19132+1035. The postulated large scale jet coming from GRS 1915+105 ends in a strong shock which gives rise to a so-called radio hotspot. This feature is identified with the non-thermal emission region at the NW edge of {\sc iras} 19132+1035. The jet material then inflates a lobe which is not detected at currently available sensitivity levels. The impact of the jet and the supersonic expansion of the lobe drive a bow shock into the surrounding ISM. The compressed and shock heated ISM in front of the jet is detected as the {\sc iras} source. We envisage a similar geometry for {\sc iras} 19124+1106. In the following we will refer to the jet leading to {\sc iras} 19124+1106 as the northern jet while the southern jet is pointing towards {\sc iras} 19132+1035.

Another prerequisite of the model is that the ejection directions of the jets remain reasonably constant.  The model requires that jet ejections propagate within the lobes created by previous jet ejections. \citet{rm99} report a possible precession angle of the small scale jets of GRS 1915+105 of about $10^{\circ}$. The two {\sc iras} regions subtend an angle of about 34" in the direction perpendicular to the jets. If this extent is determined by the precession of the jets on large scales, causing a shift of the impact region in time, then a precession angle of about $1^{\circ}$ is implied. In any case, the precession angle is small enough for the jets to remain well inside the lobes. 

The model assumes a constant energy transport rate, $Q_0$, for each jet. This transport rate is averaged over the lifetime of the jets. It is not identical to the energy transport rate estimated for the jet components observed on small scales. The mass density distribution of the ambient gas is modelled by a power law, $\rho = \rho _0 \left( r / a_0 \right) ^{-\beta}$, where $r$ is the distance from the centre of the distribution and $a_0$ is a scale length. Because GRS 1915+105 is situated in the ISM of the Galactic plane, we will in the following assume that the ambient gas has a constant density $\rho _0$ and therefore $\beta =0$. Thus we do not have to specify a value for $a_0$. KA97 show that under these circumstances the evolution of the large-scale structure created by the jet, i.e. the lobe and the bow shock in front of it, is self-similar once the jet extends beyond its characteristic length scale given by \citep[see also][]{sf91}
\begin{equation}
L_0 = \left( \frac{\rho _0^2 Q_0}{\dot{M}_0^3} \right) ^{-1/4} = \left[ \frac{ Q_0^2}{\rho _0 ^2 c^6 \left( \gamma _{\rm j} -1 \right)} \right] ^{1/4},
\label{charac}
\end{equation}
where $\dot{M}_0$ is the mass transport rate of the jet. The jet material travels with a Lorentz factor $\gamma _{\rm j}$. The second equality then follows from the assumption that the energy transported by the jet is the kinetic energy of the jet material only. In the following we estimate an upper limit for $L_0$ for the jets of GRS 1915+105 and show that the measured lengths of the two jets exceed $L_0$. Thus the model of KA97 is applicable to the case of GRS 1915+105.

In Section \ref{line} we derived an estimate for the density of the gas that has been shock heated by the impact of the jets. This gas has been compressed by the action of the bow shocks in front of the jets. For a strong shock we expect the density to be increased by a factor 4 compared to its pre-shock value $\rho_0$ \citep[e.g.][]{ll87}. Taking this into account we derive $\rho_0$ under the assumption of pure hydrogen gas (see Table \ref{modprop}).

The energy transport rate of the jets has so far only been estimated during times when jet components are detectable milliarcseconds away from the central region in the radio. During such periods, the jets are either (i) quasi-continuous with a flat radio to IR spectrum, possibly indicating self-absorbed, conical jets (e.g. \citealp{dmr00}; but see \citealp{ckc02}), or (ii) discrete, superluminal ejecta emitting optically thin synchrotron radiation are detected \citep[e.g.][]{mr94,fgmmpssw99}. During the quasi-continuous phases \citet{fp00} estimate energy transport rates between several $10^{37}$\,ergs\,s$^{-1}$ and almost $10^{42}$\,ergs\,s$^{-1}$, depending on the composition of the jet material and its velocity. Similar values are found for the discrete ejecta \citep[e.g.][]{fgmmpssw99}. Note that these estimates depend on the unknown distance $D$ of GRS 1915+105 and are valid only for limited periods of time when the jets are detectable on small scales close to the central source. As discussed above, the radio emission from the jets changes significantly over time and this radio variability may well imply variable energy input into the jets \citep[e.g.][]{kfp02}. Furthermore, the fact that GRS 1915+105 was only discovered in 1992 \citep{cbl92} in X-rays implies that the source goes through activity cycles which may also apply to the production of jets. Given the substantial evidence for the correlated behaviour of jets and accretion discs in X-ray transients, we would certainly expect this to be the case. For the properties of the large-scale structure formed by the jets only the energy transport rate of the jets averaged over timescales comparable to the fomation timescale of the large-scale structure is important. We will argue in Section \ref{transport} that this time averaged energy transport rate may be as low as $10^{36}$\,ergs\,s$^{-1}$. 

Finally, to estimate the characteristic length $L_0$ we need the Lorentz factor of the jet material. Estimating the Lorentz factor of jet flows is notoriously difficult \citep[e.g.][]{rf03}. Furthermore, the observed motion of radio `knots' in the jets may simply reflect the motion of shocks along the jets rather than the bulk motion of the jet material. Any estimate of $\gamma _{\rm j}$ will also depend on the distance to the source and the orientation of the jet to our line of sight. Here we use the conservative value of $\gamma _{\rm j}=1.1$ as a lower limit thereby maximising $L_0$.

The maximum for the characteristic jet length allowed by the plausible ranges for the relevant source parameters gives $L_0 \le 0.013\,\left( f D \right)^{1/4}$\,pc. The angular distance of the two {\sc iras} sources from GRS 1915+105, 26' 10'' for {\sc iras} 19124+1106 and 26' 24'' for {\sc iras} 19132+1035, imply jet lengths, $L_{\rm j}$, of at least $7.5\,D\,\left( \sin \theta\right)^{-1}$\,pc, where $\theta$ is the angle of  the jet axis to our line of sight. Clearly, the jets exceed their characteristic lengths by at least two orders of magnitude. Thus the model of KA97 is applicable to the jets of GRS 1915+105.

\subsection{Hydrodynamic properties of the jets}

\begin{table*}
\begin{tabular}{lll}
& Northern jet & Southern jet \\
\hline
$\dot{L}_{\rm j} / {\rm km}\,{\rm s}^{-1}$ & 23 & 23\\[0.5ex]
$L_{\rm j} / {\rm pc}$ & $7.5\,D\,\left( \sin \theta \right)^{-1}$ & $7.8\,D\,\left( \sin \theta \right)^{-1}$\\[0.5ex]
$\rho _0 / 10^{-22} \, {\rm g}\,{\rm cm}^{-3}$ & $3.1\,\left( D f \right)^{-1/2}$ & $2.0\,\left( D f \right)^{-1/2}$\\[2ex]
$t / 10^5\,{\rm years}$ & $1.9\,D\,\left( \sin \theta \right)^{-1}$ & $2.0\,D\,\left( \sin \theta \right)^{-1}$\\[0.5ex]
$Q_0 / 10^{35}\,{\rm ergs}\,{\rm s}^{-1}$ & $2.8\,D^{3/2}\,f^{-1/2}\,\left( \sin \theta \right)^{-2}$ & $2.0\,D^{3/2}\,f^{-1/2}\,\left( \sin \theta \right)^{-2}$\\[0.5ex]
$p / 10^{-10}\,{\rm ergs}\,{\rm cm}^{-3}$ & $0.31\,\left( D f \right) ^{-1/2}$ & $0.20\,\left( D f \right) ^{-1/2}$\\[2ex]
$B_{\rm min} / \mu{\rm G}$ & $22\,\left( D f \right) ^{-1/4}\,\left( k+1 \right)^{-1/2}$ & $18\,\left( D f \right) ^{-1/4}\,\left( k+1 \right)^{-1/2}$\\[0.5ex]
$S_{1.5}/ {\rm mJy}\,{\rm beam}^{-1}$ & $0.14\,\left( D f^{-7} \right)^{1/8} \left( \sin \theta \right)^{-2}\,\left( k+1 \right)^{-7/4}$ & $0.08\,\left( D f^{-7} \right)^{1/8} \left( \sin \theta \right)^{-2}\,\left( k+1 \right)^{-7/4}$\\[0.5ex]
\hline
\end{tabular}
\caption{Properties of the jets and lobes derived from observations ($\dot{L}_{\rm j}$, $L_{\rm j}$ and $\rho_0$) and from the model. $D$ is the distance of the {\sc iras} regions from us measured in kpc, $f$ is the volume filling factor with respect to a spherical emission region, $\theta$ is the angle of the jets to our line of sight, $k$ is the ratio of the energy of non-radiating particles and that of the relativistic plasma and $R$ is the aspect ratio of the lobes inflated by the jets (see text for details).}
\label{modprop}
\end{table*}

\subsubsection{Age of the jet flows}
\label{sage}

In Section \ref{line} we argued that the temperature of the hydrogen gas heated by the bow shock in front of the jets is $T=1.2\times 10^4$\,K. For a strong shock in monatomic gas we expect \citep[e.g.][]{ll87}
\begin{equation}
\frac{T}{T_{\rm i}} = \frac{5}{16} M_{\rm i}^2,
\end{equation}
where $T_{\rm i}$ is the temperature of the unshocked gas and $M_{\rm i}$ is the Mach number of the shock with respect to this material. Substituting for $M_{\rm i}$ we can solve this equation for the velocity of the bow shock which is roughly equal to the advance speed of the end of the jet driving this shock, i.e.
\begin{equation}
\dot{L}_{\rm j} \sim \sqrt{ \frac{16 k_{\rm B}}{3m_{\rm p}} T} = 23\,\mbox{km\,s$^{-1}$},
\end{equation}
where $m_{\rm p}$ is the proton mass. Clearly the bow shock will be highly supersonic with respect to the unshocked ISM, justifying our assumption of a strong shock. The kinematics of the gas in the region affected by the bow shock will be complex due to the 3-dimensional structure of the shock. It is therefore not clear whether the entire line width observed is due solely to thermal broadening. Projection of the velocity vectors of some of the shocked gas and additional turbulent motion could contribute to the observed line width implying that our estimated temperature of $1,2\times 10^4$\,K is an overestimate. In this case, we would also overestimate the velocity of the shock $\dot{L}_{\rm j}$. In the following we will use the value of $23$\,km\,s$^{-1}$, but the reader is reminded that this may well be a upper limit for the shock velocity rather than its exact value. 

From KA97 we take the result that
\begin{equation}
L_{\rm j} = c_1 \left( \frac{Q_0}{\rho _0} \right)^{1/5} t^{3/5}
\label{length}
\end{equation}
and therefore 
\begin{equation}
\dot{L}_{\rm j} = \frac{3 L_{\rm j}}{5 t},
\label{age}
\end{equation}
where $t$ is the age of the jet. $c_1$ is a dimensionless constant which depends on the thermodynamical properties of the jet material. It also depends on the aspect ratio, $R$, of the lobe inflated by the jet, i.e. $R$ is the length of the lobe divided by its width. With the measured lengths given in Table \ref{modprop}, we find the ages of the jets from equation (\ref{age}) to be of order 10$^6$\,years (see Table \ref{modprop}). 

We currently do not understand the connection between the processes in the accretion disc and the formation of jets in binary systems. However, GRS 1915+105 is currently in a state with low accretion rates. \citet{bb02} estimate that this stage can last for about $10^7$\,years. If the production of jets is associated with the low accretion state of this system, then our age estimate for the jets is consistent with this finding.

\subsubsection{Energy transport rate of the jets and the pressure in the lobes}
\label{transport}

With these results we can determine the energy transport rate, $Q_0$, of the jets from equation (\ref{length}). The constant $c_1$ is unknown because the lobes have not been detected and therefore their aspect ratios are unknown. However, for extragalactic jet sources $R\gtappeq 2$ is found with very few exceptions \citep[e.g.][]{lw84,lms89}. As a limiting case we adopt $R \sim 2$ and thus $c_1=2$ in the following. The derived values for $Q_0$ for this assumption are given in Table \ref{modprop}. Note that $Q_0$ depends strongly on the value of $c_1$, but any value $R>2$ will result in a lower value of $Q_0$.

For the usually assumed distance of GRS 1915+105 of $D=11$\,kpc and a viewing angle of roughly $\theta =70^{\circ}$ the energy transport rate of the jets is comparable to the lower end of the range of $Q_0$ estimated at times when the small scale jets are detectable (see Section \ref{basic}). This result  is consistent with the fact that our energy transport rate is averaged over the lifetime of the jets, i.e. over timescales of order $10^5$ to $10^6$\,years, while the small scale jets used for other estimates have much shorter lifetimes. In this context it is interesting to note that our value for $Q_0$ is not much lower than the energy transport rate inferred for the small scale jets. This strongly suggests that the active phases during which GRS 1915+105 produces jets must be at least comparable in length to the passive phases when jet production is switched off. On the other hand, episodes of very powerful jet ejections are probably rare as they do not seem to dominate the time averaged energy budget of the jets.

In the model of KA97 the pressure inside the lobes is given by
\begin{equation}
p = 0.0675 \frac{c_1^{10/3}}{R^2} \left( \frac{\rho _0 Q_0^2}{L_{\rm j}^4} \right)^{1/3}.
\end{equation}
For $R=2$ we find that $c_1^{10/3} / R^2 \sim 1$ and the resulting lobe pressures are given in Table \ref{modprop}. For the density of the unshocked ISM derived above and an assumed temperature of this gas of 50\,K, we find a pressure of roughly $1.1\times10^{-12}\,\left( D f \right) ^{-1/2}$\,ergs\,cm$^{-3}$. The lobes will clearly be overpressured with respect to the unperturbed ISM, as required by the model. 

\subsection{Emission properties of the model}

\subsubsection{Radio emission from the lobes}

In the case of extragalatic jet sources a substantial fraction of the energy transported by the jets gives rise to a population of relativistic electrons and a magnetic field in the lobes. The synchrotron emission from this relativistic plasma is easily detected in many such objects in the radio. It is usually assumed that the relativistic electrons are accelerated by the first order Fermi process at the strong shocks at the ends of the jets. Even if they are transported by the jets all the way from the central source \citep{gwb01}, there is no {\em a priori\/} reason to assume that the lobes of microquasar jets should not also produce radio synchrotron emission. However, searches for such lobes have been unsuccessful for GRS 1915+105 (RM98). In the following we will derive an upper limit for the surface brightness of the expected radio emission from the lobes and show that it is well below the sensitivity of the radio maps produced to date.

Assuming that the magnetic field is tangled on scales much smaller than the lengths of the jets, we will have three contributions to the overall pressure in the lobe derived in the previous section: The pressure due to the magnetic field, the pressure of the relativistic electrons and that of any other particles in the lobes which do not emit synchrotron radiation, e.g. protons. As in Section \ref{nontherm} the parameter $k$ is the ratio of the energy density of the non-radiating particles and the sum of the energy densities of the relativistic electrons and the magnetic field. Since we want to derive an upper limit for the radio surface brightness, we assume that the usual conditions of minimum energy hold in the lobe. In this case, the maximum synchrotron luminosity is obtained from a given amount of energy \citep[e.g.][]{ml94}.   We can neglect any energy losses of the relativistic electrons due to radiation processes because the age of the source derived above is well below the timescale for radiative losses. Solving equation (\ref{minpres}) for the strength of the magnetic field we find $B_{\rm min}$; see Table \ref{modprop}.

We can now solve equation (\ref{bmin}) for the monochromatic luminosity $L_{\nu}$ of the lobes predicted by the model. However, because of the lack of direct detections of emission from the lobes we have to guess the values of $\nu_{\rm min}$, $\nu_{\rm max}$ and $\alpha$. The cut-offs of the spectrum do not greatly influence our result and we set them to $\nu_{\rm min} = 10^7$\,Hz and $\nu_{\rm max} = 10^{11}$\,Hz as in Section \ref{nontherm}. We will argue further down that the non-thermal emission region attached to {\sc iras} 19132+1035 is the strong shock of the southern jet. The synchrotron emission from this region has a spectral index of $\alpha =-0.8$. Since in our picture the relativistic plasma passing through this shock inflates the southern lobe, it is reasonable to use the same spectral index to derive $L_{\nu}$ for the lobes.

A single telescope beam of angular diameter $\delta$ roughly contains the flux emitted by a source volume given by 
\begin{equation}
V=\pi \left( D \delta / 2 \right)^2 l,
\end{equation}
with 
\begin{equation}
l=L_{\rm j} / \left( R \sin \theta \right)
\end{equation}
the intersection of maximum length of our line of sight with the lobe. Note here that $R$ is the real aspect ratio of the lobe. For a viewing angle $\theta \ne 90^{\circ}$ this differs from the observed aspect ratio by a factor $\sin \theta$. Solving equation (\ref{bmin}) for $L_{\nu}$ and converting this to flux density, we can now determine the upper limit to the surface brightness of the lobe, $S_{\nu}$, as predicted by the model. The observational limits are set by the data of RM98 with an approximate noise 1-$\sigma$ rms noise level of 0.2\,mJy per beam with $\delta \sim 4$" at $\nu = 1.5$\,GHz. The limits are the same for the other, higher observing frequencies. Given the expected power law shape of the spectrum we use this lowest observing frequency, i.e. 1.5\,GHz, in our estimate because it provides the tightest constraint. The results are summarised in Table \ref{modprop}. Note that for thinner lobes with $R>2$ these surface brightness limits decrease further. Any deviation from the conditions of minimum energy will further reduce the surface brightness. Unless $f \ll 1$ and/or our viewing angle to the jets is small, this upper limit is below the detection limit of current radio observations. This conclusion is almost independent of the distance of GRS 1915+105. Therefore it is not surprising that the synchrotron radio emission of the lobes has not been detected to date. 

Recently, \citet{sh02} \citep[see also][]{sh03} has developed another approach for estimating the surface brightness of the possible radio lobes of microquasars similar to the one presented here. Although there are difference in details, the models are sufficiently similar to warrant a comparison. \citet{sh02} concluded that to explain the non-detection of any lobe emission from GRS 1915+105 it must be located in an environment with an unusually low density of below $10^{-3}$\,cm$^{-3}$. This apparent discrepancy between the models is explained by their assumption of an energy transport rate of $10^{46}$\,ergs\,s$^{-1}$ for the jets. Although appropriate for the jets during major outbursts with powerful small scale jets, our model suggests that this is an overestimate for the time averaged energy transport rate. Using our value of roughly $10^{37}$\,ergs\,s$^{-1}$ we find that the predictions of \citet{sh02} agree with ours, namely that the lobes should not be detectable in current radio maps.

\subsubsection{Radio emission from the hotspots}

The sites of the strong shocks at the end of the jets of extragalactic jets are often referred to as radio hotspots or simply hotspots. It is usually assumed that the relativistic electrons giving rise to the synchrotron emission of the radio lobes of these objects are accelerated in these hotspots \citep[but see][for an alternative view]{gwb01}. Because of the availability of freshly accelerated relativistic electrons and the enhancement of the magnetic field in these high pressure regions, the hotspots are considerably brighter than the radio lobes. If our model is correct, then the jets of GRS 1915+105 should also end in strong shocks and therefore produce radio hotspots which should be easier to detect than the lobes. The non-thermal emission region at the northwestern edge of {\sc iras} 19132+1035 could be such a hotspot. The model we use in previous sections does not make direct predictions for the radio emission from hotspots. However, there is some circumstantial evidence supporting the identification proposed here.

The non-thermal feature near {\sc iras} 19132+1035 is clearly elongated in the radio maps. The longer axis points from {\sc iras} 19132+1035 back towards the position of GRS 1915+105. Both RM98 and CRM01 quote this property as the only tentative evidence that the two {\sc iras} regions are connected to GRS 1915+105 by jets. 

The pressure inside the feature derived in Section \ref{nontherm} is close to the pressure of the shocked ISM (see Tables \ref{gasprop} and \ref{nthprop}). The model presented here is based on the assumption that the pressure at the end of the jets is balanced by the pressure of the surrounding gas compressed by the bow shock. Our findings are consistent with this pressure equilibrium and therefore imply that the pressure in the non-thermal feature is dominated by the contribution of the relativistic gas, i.e. $k$ is small. A significant contribution to the total energy density by non-radiating particles would imply that the feature is overpressured with respect to the shocked ISM. The approximate equality of the pressure of the non-thermal feature and the shocked ISM also makes it unlikely that the relativistic plasma in the feature is far from the conditions of minimum energy. 

For an approximately cylindrical lobe we expect the ratio of the pressure in the hotspot and that in the lobe to be roughly $4R^2$ (KA97). Identifying the non-thermal feature with the hotspot of the southern jet we find from the results of Section \ref{nontherm} and \ref{transport}
\begin{equation}
R \sim 2.1\,D^{-1/28}\,\left( \sin \theta \right)^{2/7} \left( k +1 \right)^{1/2}\,f^{1/4}.
\end{equation}
This aspect ratio is comparable to those found for lobes of extragalactic jet sources \citep[e.g.][]{lw84,lms89} and justifies our assumption of $R\sim 2$.

At 1.5\,GHz the predicted surface brightness of the non-thermal feature is at least a factor 250 greater than that of the undetected southern lobe. However, the scale of this difference is not surprising as the surface brightness of hotspots of extragalactic jet sources can be orders of magnitude greater than that of their lobes. 

There is no equivalent non-thermal emission region associated with the northern jet. Given that the small scale jets are known to be transient, it is possible that the jets on larger scales and therefore the hotspots at their ends are not continuously fed with relativistic plasma.  Any relativistic plasma transported by the southern jet into the non-thermal feature would pass through the feature along the long axis within a time 
\begin{equation}
t_{\rm cross} = 100\,D\,\left( v \sin \theta \right) ^{-1}\,\mbox{days},
\end{equation}
where $v$ is the velocity of the plasma in units of the speed of light. Clearly the hotspots could therefore be transient phenomena as well which disperse on timescales of several months to a few years. Subsequently they may be undetectable for similar timescales until more plasma arrives from the central source via the jets. 

In this context it is interesting to note that the shock tracer SiO was only detected in {\sc iras} 19132+1035 which also contains the non-thermal emission feature (see Section \ref{molecular}). If jet material is currently arriving at the end of the jet pointing towards {\sc iras} 19132+1035, then this material should form a hotspot, which we identify with the non-thermal feature, and it should drive a bow shock into the ISM giving rise to the SiO line emission. For {\sc iras} 19124+1106 no jet material is currently arriving at the end of the jet and so no hotspot and no SiO lines are detected. 

\subsubsection{Dust emission}

In our model the bow shock in front of the jets and lobes will heat the ISM and any dust mixed in with it. This process will give rise to at least some of the IR emission observed in the two {\sc iras} regions. One obvious question arising from this scenario is as to whether it is
realistic to find dust cospatial with hot gas ($T \sim
10^4$\,K), and in the harsh environment of the bow shocks driven by the jets.  However, the
survival of dust grains has been studied extensively, such as in
regions of jet-gas interactions in active galaxies
\citep[e.g.][]{VM01}, and in supernova remnants
\citep[e.g.][]{Mouri00}.  In the case of slow-moving shocks, such as
here, the destruction of dust grains is found to be a comparatively
inefficient process \cite{DeYoung98}, and we would not expect them to
be destroyed by the jets of GRS 1915+105. 

The observed IR luminosities are very high and the emission regions seem to extend well beyond the confines of the radio bremsstrahlung emission regions for both {\sc iras} sources. These observations may imply an additional mechanism for the heating of the dust on large scales, away from the immediate impact sites of the jets. Also, the energy transfer from hot gas to dust with the addition of radiative losses is a complicated process and modelling this is beyond the scope of this paper. However, the emission peak at 15\,$\mu$m coincides with the peak of the bremsstrahlung emission in both {\sc iras} regions (see Figure 6 of CRM01) suggesting a close link between the jet impact sites and the heating of the dust. 

\section{The distance to GRS 1915+105, the velocity of the jets and their orientation}
\label{dis}

\subsection{Distance estimates}
\label{distance}

\subsubsection{Distance implied by the model}

If the two {\sc iras} sources are indeed the endpoints of the jets of GRS 1915+105, then we would expect GRS 1915+105 and both {\sc iras} sources to all lie at a similar distance from us. RM98 observed the centre of the H92$\alpha$ recombination line to be shifted from its restframe frequency by 57.3\,km\,s$^{-1}$ for {\sc iras} 19124+1106 and by 75.7\,km\,s$^{-1}$ for {\sc iras} 19132+1035. If these shifts arise from the difference in the Galactic orbital velocity of these objects with respect to us, then they imply distances of between 6\,kpc and 7.5\,kpc. The observed line shifts are essentially confirmed by the millimeter molecular lines detected by CRM01, but given the narrower lines and therefore smaller uncertainties in these observations, their line shift of about 67\,km\,s$^{-1}$ for {\sc iras} 19132+1035 is more likely to be correct. 

In our model an additional velocity component is added to the gas in the thermal bremsstrahlung emission region by the action of the bow shock accelerating the gas in front of the jet ends. Before we can determine the distance of the {\sc iras} regions from us by using the Galactic rotation curve, we need to correct for this effect. Gas at rest with respect to the mean rotation of the Galaxy local to the gas will show a line shift corresponding to the velocity $v_{\rm LSR}$, where `LSR' stands for Local Standard of Rest. If this gas is pushed by the jets, then the gas in front of the jet at angle $\theta$ to our line of sight pointing preferentially towards us, the approaching jet, will have a velocity with respect to us equal to 
\begin{equation}
v_{\rm ap} = v_{\rm LSR,ap}+\dot{L}_{\rm j} \cos \theta. 
\label{app}
\end{equation}
By symmetry, the gas in front of the jet pointing preferentially away from us, the receding jet, will have a velocity equal to
\begin{equation}
v_{\rm rec} = v_{\rm LSR,rec}-\dot{L}_{\rm j} \cos \theta. 
\label{rec}
\end{equation}
Eliminating the term $\dot{L}_{\rm j} \cos \theta$ between these two equations yields an expression relating the velocities $v_{\rm LSR,ap}$ and $v_{\rm LSR,rec}$. Both depend only on the known positions on the sky of the two gas clouds, the distance of the gas clouds from us and on the adopted rotation curve of the Galaxy. From observations of the small scale jets we find that the southern jet associated with {\sc iras} 19132+1035 is the approaching jet and so we set $v_{\rm ap}=67$\,km\,s$^{-1}$ and $v_{\rm rec} = 57.3$\,km\,s$^{-1}$. Using the rotation curve of \citet{bb93} we then find two possible solutions, $D_1$ and $D_2$, for the distance of GRS 1915+105. Either we have $v_{\rm LSR,ap} = 61.6$\,km\,s$^{-1}$ and $v_{\rm LSR,rec} = 62.8$\,km\,s$^{-1}$ implying $D_1=5.4$\,kpc, or $v_{\rm LSR,ap} = 61.4$\,km\,s$^{-1}$ and $v_{\rm LSR,rec} = 63.0$\,km\,s$^{-1}$ with $D_2=6.5$\,kpc. Even after correcting for the effects of the bow shock the inferred rotation velocities of the gas clouds are comparable to the maximum rotation velocity close to the tangent point in the Galaxy at a distance of 6\,kpc. This explains the small difference between $D_1$ and $D_2$. 

We derive the distance estimates above under the implicit assumption that the correction factors in equations (\ref{app}) and (\ref{rec}) are equal in magnitude and opposite in sign. The complex gas flow in the vicinity of the bow shocks in front of the jets makes it impossible to find a definite value for the velocity of the bow shock, $\dot{L}_{\rm j}$ and thus for the `real' correction factors for $v_{\rm ap}$ and $v_{\rm rec}$. Therefore there is an uncertainty associated with the distance estimates $D_1$ and $D_2$ and in particular with the orientation angle $\theta \sim 80^{\circ}$ implied by equations (\ref{app}) and (\ref{rec}). However, as shown by RM98 the large measured line shifts of at least $57.3$\,km\,s$^{-1}$ rule out a distance of the {\sc iras} regions of greater than 7.4\,kpc. It is therefore not surprising that the distances estimated from the Galactic rotation curve here are considerably smaller than the distance usually assumed for GRS 1915+105.

\subsubsection{Other distance estimates}

As mentioned in the previous section, the distance estimates to the two {\sc iras} sources do not allow them to be located far beyond the tangent point at 6\,kpc. Attempts have been made to measure the distance to GRS 1915+105 itself. \citet{dgr00} used the Galactic H{\sc i} absorption observed during ejections of radio emitting small scale jets. They conclude that GRS 1915+105 must be at a distance beyond the tangent point at 6.1\,kpc but closer than 12.2\,kpc. By comparison with a nearby H{\sc ii} region they suggest a distance of greater than 9\,kpc. \citet{cc04} study the optical extinction towards GRS 1915+105 for a distance estimate. They derive a very similar lower limit to the distance of 6\,kpc, but also show that the argument of  \citet{dgr00} for $D>9$\,kpc does not apply. Both studies agree that the source should be beyond the tangent point at 6\,kpc. This suggests that if the {\sc iras} sources are in fact associated with GRS 1915+105, then our second estimate $D_2 = 6.5$\,kpc is likely to be more accurate than $D_1$. 

The observed motion of components in the small scale jets also provides a constraint on the distance. As these components cannot move intrinsically faster than the speed of light, \citet{fgmmpssw99} show that GRS 1915+105 cannot be further away than 11.2\,kpc. Unfortunately, this method does not provide us with a lower limit for $D$. 

Finally, two attempts have been made to directly measure the motion of GRS 1915+105 in the Galaxy and thus determine $D$ using the Galactic rotation curve. \citet{dmr00} determined the proper motion of the radio counterpart of GRS 1915+105. They found a velocity of $5.8\pm1.5$\,mas\,year$^{-1}$ which is consistent with the limits derived from absorption measurements quoted above, i.e. $6\,{\rm kpc} \le D \le 12\,{\rm kpc}$. The determination of the system period by \citet{gcm01} provides an estimate of the line of sight velocity of GRS 1915+105. The off-set of the measured radial velocity curve determines the systemic velocity with respect to the observer. From the Galactic rotation curve they then find a distance of $12.1\pm0.8$\,kpc. Although this measurement seems to rule out that GRS 1915+105 and the two {\sc iras} regions are at a common distance of 6\,kpc, it should be borne in mind that the main aim of \citet{gcm01} was the determination of the orbital period of the binary system. However, the orbital period can be obtained independent of the absolute calibration of the observed spectra and systematic effects may well have altered the determination of the systemic velocity. An independent confirmation of the off-set of the radial velocity curve and thus the systemic velocity would clarify this issue. 

The model presented here does not directly depend on the distance of GRS 1915+105. In other words, our results remain valid even if the source is located at a distance of 11 or 12\,kpc from us. However, in this case the large velocity of the two {\sc iras} sources along our line of sight cannot be explained by Galactic rotation. In principle the extra velocity could come from a kick the system received during the formation of the black hole. This would then mean that the entire structure, i.e. GRS 1915+105 and both {\sc iras} regions, are moving together with the kick velocity. There are several problems with this explanation. Firstly, we would still expect to see a systemic velocity of GRS 1915+105 itself comparable to that of the {\sc iras} sources. Secondly, the kick velocity would need to have a component of order 50\,km\,s$^{-1}$ along our line of sight. The ends of the jets only advance with a velocity of 23\,km\,s$^{-1}$ and so the formation of a stable jet channel and the lobe structure would be impossible. In fact, \citet{bb02} suggest that GRS 1915+105 did not experience a kick associated with the formation of the black hole.

In summary we emphasize that the model presented here is only valid if GRS 1915+105 and both {\sc iras} regions are located at the same distance from us. The {\sc iras} regions are almost certainly close to the tangent point at 6\,kpc. Further observations to ascertain the distance of GRS 1915+105 are needed and we predict that such studies will find the source to be located at 6.5\,kpc from us.

\subsection{Jet velocity and orientation}
\label{veloc}

As mentioned above, studies of the observed motion of components of the small scale jets provide us only with an upper limit on the distance of the source. Furthermore, without a secure distance measurement we cannot determine the velocity of the jet material, $v_{\rm j}$, separately from the jet orientation. \citet{fgmmpssw99} derive an expression for the jet orientation to our line of sight depending on $D$ (their equation 4). From this expression, it is straightforward to infer the velocity of the jet material. Using our distance estimates $D_1$ and $D_2$ from above based on the Galactic rotation curve, we find for their radio data $\theta _1=47^{\circ}$ and $v_{\rm j 1} = 0.61$\,c for $D_1$ and $\theta _2 = 53^{\circ}$ and $v_{\rm j2} = 0.68$\,c for $D_2$. Both velocities imply only mildly relativistic fluid flow in the jets with Lorentz factors around 1.3. 

\citet{rf03} points out that for significantly relativistic jet speeds we are biased to observe jets at their maximum distance from us. If the jets of GRS 1915+105 were highly relativistic, this would argue for a distance close to 11\,kpc for GRS 1915+105. However, for the distance required by our model the jet flow is only mildly relativistic and so the argument does not apply.

The velocity of the jet and its orientation to our line of sight inferred for a distance of 11\,kpc by the method of \citet{fgmmpssw99} are $v_{\rm j}=0.98$\,c and $\theta = 66^{\circ}$. Any emission line produced by the jet material at a rest-frame frequency $\nu_0$ would therefore be Doppler-shifted to $\nu_{\rm app} = 0.33 \nu_0$ for the approaching jet and to $\nu_{\rm rec} = 0.14 \nu_0$ for the receding jet. These results are clearly different from the predictions of our model of $\nu_{\rm app} = 1.36 \nu_0$, $\nu_{\rm rec} = 0.56 \nu_0$ in the case of $D_1$ and $\nu_{\rm app} = 1.24 \nu_0$, $\nu_{\rm rec} = 0.52 \nu_0$ for $D_2$. The question over the correct distance to GRS 1915+105 could therefore be resolved if a pair of emission lines from the small-scale jets could be detected. In practice the large extinction towards the source may make such an observation impractical.

Note, that the orientation angle estimated here does not agree with that implied by equations (\ref{app}) and (\ref{rec}). As mentioned in the previous section, the constraints on $\theta$ from the Galactic rotation curve are not very strong given the uncertainties for $\dot{L}_{\rm j}$ and the appropriate correction factor for the rotation velocities of the {\sc iras} regions. 

\section{Conclusions}
\label{conc}

In this paper we have applied the model of KA97 for the large scale structure of jets in extragalatic radio sources to the microquasar GRS 1915+105. The jets emerging from the central source end in strong shocks where they impact on the surrounding ISM. The jet material subsequently inflates overpressured lobes which expand at supersonic velocities. The ISM surrounding the lobes is compressed by the resulting bow shock. This shock is strongest in front of the ends of the jets. 

Although no radio emission from the lobes themselves is observed, we identify the two {\sc iras} regions {\sc iras} 19124+1106 and {\sc iras} 19132+1035 as the jet impact regions. The impact regions are almost perfectly symmetric about the position of GRS 1915+105 and they lie in the directions that the small scale jets of this source point to. The radio emission from these impact regions has a flat spectrum which we interpret as bremsstrahlung from the shock compressed and heated ISM. From the flux density of this emission we derive the density of the ISM, while the width of the H92$\alpha$ recombination line is used to infer the velocity of the bow shock. Together with the observed length of the jets, the density of the shocked gas and the shock velocity allow us to derive the age and energy transport rate of the jets. The latter is lower than the jet power estimated for the large, discrete ejection events observed on small scales. However, these events are short-lived and the energy delivered to the lobes by the jets is governed by the energy transport rate averaged over the lifetime of the jets, of order $10^6$\,years.

From the pressure predicted by the model we derive an upper limit for the radio surface brightness of the lobes under the assumption of minimum energy conditions. We show that current radio observations have not gone deep enough to reveal the lobes. We do not confirm claims of an unusually low density of the source environment based on this non-detection \citep{sh02,sh03}. Instead, the low radio surface brightness is caused by the intermittent jet activity that leads to a lower average energy transport rate compared to major ejection events observed on small scales.

Further support for our interpretation comes from the observation of a non-thermal, radio emitting region near to {\sc iras} 19132+1035. This feature is elongated and appears to point back towards GRS 1915+105. We identify this feature as the strong shock at the jet end or `radio hotspot' of the southern jet of GRS 1915+105. The pressure inside this region derived from its radio emission and projected size is consistent with this idea. The absence of a similar feature for the northern jet pointing towards {\sc iras} 19124+1106 can be explained by the transient nature of radio hotspots and the intermittency of the jet activity. 

The observations of molecular lines and dust emission co-spatial with the {\sc iras} regions is consistent with our interpretation.

The model requires that GRS 1915+105 and the two {\sc iras} regions are located at the same distance from us. Emission line shifts place the {\sc iras} regions at a distance of about 6.5\,kpc. This value is consistent with distance estimates for GRS 1915+105 from absorption studies in H{\sc i} and the optical waveband as well as with the observations of motion in the small scale jets. The systemic velocity of GRS 1915+105 was derived as a byproduct of the determination of the period of the binary system \citep{gcm01}. This measurement suggests a distance of 12\,kpc to the source, also consistent with the results of the other methods, but in conflict with our model.  

Our model is a self-consistent interpretation of all currently available data on the large scale jet structure of GRS 1915+105 with the possible exception of the distance of this source from us as determined by \citet{gcm01}. However, the correct value for the distance is still uncertain and will require further observations. One possibility to pin down the correct distance would be detection of an emission line pair from the approaching and receding jets. The Doppler shifts predicted for such a line pair are significantly different for different distances of the source. Our model predicts that such studies will find a distance of about 6.5\,kpc. The model further predicts that the radio hotspots of the jets should be located close to the two {\sc iras} regions. Due to the intermittency of the jet activity and their transient nature, follow-up radio observations of the {\sc iras} regions should show significant change of the morphology of the southern non-thermal hotspot and may very well reveal a similar emission region in the vicinity of {\sc iras} 19124+1106. Finally, the radio lobes themselves should be detectable in deeper radio observations, possibly at lower observing frequencies. 

The model presented here is a first step towards an understanding of the impact of microquasar jets on their environment. The energy transported by the jets is found to give rise to observable features in the ISM. Further studies of objects other than GRS 1915+105 are needed to confirm this picture. The large scale structure to which the jets of microquasars give rise is comparable to that caused by extragalactic jets. Thus the way in which the jets interact with their environments is another similarity between the two object classes.

\acknowledgments{The authors thank K. Belczynski, R.P. Fender, C. Knigge, U. Lisenfeld and G. Pavlovski for useful discussions. We also thank the anonymous referee for helpful comments. JLS is supported by an NSF Astronomy and Astrophysics Fellowship under award AST-0302055.}

\bibliography{crk}
\bibliographystyle{mn2e}

\end{document}